\documentclass[epj]{svjour}
\usepackage{graphics}
\usepackage{epsfig}
\usepackage{amssymb}
\usepackage{amsbsy}

\begin{document}
\title{First-forbidden $\mathbf{\beta}$-decay rates, energy rates of
$\beta$-delayed neutrons and probability of $\beta$-delayed neutron
emissions for neutron-rich nickel isotopes.}
\author{Jameel-Un Nabi  \inst{1} \thanks{Corresponding author email: jameel@giki.edu.pk}, Necla \c{C}akmak \inst{2} and Zafar
Iftikhar \inst{1}} \institute{
  \inst{1} Faculty of Engineering
Sciences, GIK Institute of Engineering Sciences and Technology,
Topi 23640, Swabi, Khyber Pakhtunkhwa, Pakistan\\
  \inst{2} Department of Physics, Karab\"uk University, 78050, Karab\"uk, Turkey}

\abstract{First-forbidden (FF) transitions can play an important
role in decreasing the calculated half-lives specially in
environments where allowed Gamow-Teller (GT) transitions are
unfavored. Of special mention is the case of neutron-rich nuclei
where, due to phase-space amplification, FF transitions are much
favored. We calculate the allowed GT transitions in various pn-QRPA
models for even-even neutron-rich isotopes of nickel. Here we also
study the effect of deformation on the calculated GT strengths. The
FF transitions for even-even neutron-rich isotopes of nickel are
calculated assuming the nuclei to be spherical. Later we take into
account deformation of nuclei and calculate GT + unique FF
transitions, stellar $\beta$-decay rates, energy rate of
$\beta$-delayed neutrons and probability of $\beta$-delayed neutron
emissions. The calculated half-lives are in excellent agreement with
measured ones and might contribute in speeding-up of the $r$-matter
flow.}
\authorrunning{Nabi, \c{C}akmak and Iftikhar}
\titlerunning{FF $\beta$-decay rates for neutron-rich Ni isotopes}
\maketitle
\onecolumn
\section{Introduction}
Exotic nuclei exhibiting high isospin values and located far from
stability line have gained amplified interest. The reason for this
is twofold. Unprecedented features are predicted with recent
theoretical development when moving towards the neutron drip line
and away from the valley of stability \cite{Dob94}. Changing of the
traditional shell gaps and magic numbers, spin-orbit interaction
weakening and dilute neutron matter are some of the examples of
these features. The study of nuclei structure and its understanding
when put under extreme conditions can shed new light on the
effective nucleon-nucleon interaction choice in the nuclear medium.
Following to this, the so called neutron-rich exotic nuclei are
believed to play a crucial role in explosive nucleosynthesis
phenomenon like the different $r$-processes in supernovae events
\cite{Kra93}.

It is commonly accepted that $r$-process occurs in an explosive
environment of relatively high temperatures (T $\thickapprox$
10$^{9}$ K) and very high neutron densities ($>$10$^{20}$ cm$^{-3}$)
\cite{Kra93,Bur57,Cam57,Cow91,Woo94}. Neutron captures, under such
conditions, are observed much faster than competing $\beta$-decays
and the $r$-process path in the nuclear chart proceeds through a
chain of extremely neutron-rich nuclei with approximately constant
and relatively low neutron separation energies (S$_{n}$ $\lesssim$ 3
MeV). They form a chain through which the $r$-process path appears
to proceed in the nuclear chart. The neutron separation energies
show discontinuities at the magic numbers N = 50, 82, and 126 due to
the relatively stronger binding of nuclei with magic neutron
numbers. The $r$-matter flow, in consequence, slows down when it
approaches these magic neutron nuclei and here it waits for several
$\beta$-decays (which are also longer than for other nuclei on the
$r$-process path) to occur before further neutron captures are
possible, carrying the mass flow to heavier nuclei. Thus matter is
accumulated at these $r$-process waiting points associated with the
neutron numbers N = 50, 82, and 126, leading to the well-known peaks
in the observed $r$-process abundance distribution.

Practically in all stellar processes, e.g., massive stars
hydrostatic burning, pre-supernova evolution of massive stars and
nucleosynthesis ($s$-, $p$-, $r$-, $rp$-) processes, the weak
interaction rates are the important ingredients and play a crucial
role \cite{Bur57}. Stellar weak interaction processes, for densities
$\rho \lesssim$ 10$^{11}$ g/cm$^{3}$, are dominated by Gamow-Teller
(GT) and also by Fermi transitions if applicable. The contribution
of forbidden transitions is observed sizeable for nuclei lying in
the vicinity of $\beta$ stability line for density $\rho\geq$
10$^{11}$ g/cm$^{3}$ and electron chemical potential of the order of
30 MeV or more \cite{Coo84}. However recent studies have shown the
importance of forbidden transitions also at orders of magnitude
lower densities \cite{Nab13,Zhi13}. The QRPA studies based on the
Fayans energy functional has been extended by Borzov recently for a
consistent treatment of allowed and first-forbidden (FF)
contributions to $r$-process half-lives \cite{Bor06}. A significant
reduction in the half-lives of N = 126 is seen while these
calculations find that forbidden contributions give only a small
correction to the half-lives of the N = 50 and N = 82.  The
correlations among nucleons are expected to not only affect the
half-lives, but to give a reliable description of the detailed
allowed and forbidden strength functions. This description is indeed
needed to estimate the energy rate of neutron from daughter nuclei
and probabilities for beta-delayed neutron emission rates which are
known to be important to describe the decay of the $r$-process
nuclei towards stability after freeze-out.

The $\beta$-decay properties, under terrestrial conditions, of
allowed weak interaction and U1F \cite{Hom96} led to a better
understanding of the $r$-process. Authors in \cite{Hom96} showed
that for near-stable and near-magic nuclei a large contribution to
the total transition probability came from U1F transitions. To
describe the isotopic dependence of the $\beta$-decay
characteristics the allowed $\beta$-decay approximation alone is not
sufficient \cite{Bor06}, specially for the nuclei crossing the
closed N and Z shells for which forbidden transitions give a
dominant contribution to the total half-life (specially for N $>$ 50
in $^{78}$Ni region). A large-scale shell-model calculation of the
half-lives, including first-forbidden contributions, was also
performed for $r$-process waiting-point nuclei \cite{Zhi13}. Since
the weak interaction rates are of decisive importance in the domains
of high temperature and density therefore there was a need to
perform these calculations under stellar conditions. The microscopic
calculations of allowed GT and unique first-forbidden (U1F) rates
for nickel isotopes in stellar environment were performed recently
using the pn-QRPA model \cite{Nab13}. This study suggested to also
incorporate rank 0 and rank 1 operators for a full coverage of FF
transitions and much better comparison with measured half-lives.

The pn-QRPA model was developed by Halbleib and Sorensen
\cite{Hal67} by generalizing the usual RPA to describe
charge-changing transitions. A microscopic approach based on the
proton neutron quasi-particle random phase approximation (pn-QRPA),
have so far been successfully used in studies of nuclear
$\beta$-decay properties of stellar weak-interaction mediated rates
(see \cite{Nab99,Nab99a,Nab04}). The pn-QRPA model allows a
state-by-state evaluation of the weak rates by summing over
Boltzmann-weighted, microscopically determined GT strengths for all
parent excited states. Construction of a quasi-particle basis is
first performed in this model with a pairing interaction, and then
the RPA equation is solved with GT residual interaction.

This paper can be broadly categorized into two different
calculations. Both calculations employ  pn-QRPA methods  using two
different potentials. The first calculation deals with the allowed
GT strength  of even-even spherical and deformed isotopes of
neutron-rich nickel isotopes using the Woods-Saxon potential. Here
we use three different versions of the pn-QRPA model and tag them as
pn-QRPA(WS-SSM), pn-QRPA(WS-SPM) and pn-QRPA(WS-DSM) models for
spherical and deformed cases, respectively.  The FF contributions
were calculated using only the pn-QRPA(SSM) model. The second
pn-QRPA calculation employs a deformed Nilsson potential and is
referred to as pn-QRPA(N) model. All models would be introduced in
the next section. The pn-QRPA(N) model was used to calculate GT+U1F
transitions, $\beta$-decay and positron capture rates, energy rate
of emitted neutrons from daughter nuclei and probability of
$\beta$-delayed neutron emissions. All pn-QRPA(N) calculations were
further performed in stellar environment.

The paper is organized in four sections. Section~2 briefly describes
the formalism of the various pn-QRPA models used in this paper. In
Section~3, we discuss the results of allowed GT and FF strength
distributions, phase space calculations, half-lives, stellar
$\beta$-decay and positron capture rates, energy rate of neutrons
and probability of $\beta$-delayed neutron emissions for the
neutron-rich nickel isotopes. Section~4 finally concludes our
findings.

\section{Formalism}
As discussed earlier, this paper can be broadly divided into two
different set of calculations using pn-QRPA models with different
single-particle potentials. In this section we briefly describe the
formalism to perform the respective calculations using the different
models.

\subsection{The pn-QRPA(WS) model}
Allowed beta decay half-lives have been calculated using deformed
schematic model (DSM), spherical schematic model (SSM) and spherical
Pyatov's method (SPM) within the framework of pn-QRPA(WS) method.
The Woods-Saxon potential with Chepurnov parametrization has been
used as a mean field basis in numerical calculations. The
eigenvalues and eigenfunctions of the Hamiltonian with separable
residual GT effective interactions in particle-hole (ph) channel
were solved within the framework of pn-QRPA model.

We constructed a quasi-particle basis described by a Bogoliubov
transformation and solved the RPA equation with a schematic residual
GT interaction
\begin{eqnarray}
a_{j_{p}m_{p}}=U_{j_{p}}\alpha_{j_{p}m_{p}}+(-1)^{j_{p}-m_{p}}V_{j_{p}}\alpha^{+}_{j_{p}-m_{p}} \nonumber
\end{eqnarray}
\begin{eqnarray}
a^{+}_{j_{n}m_{n}}=U_{j_{n}}\alpha^{+}_{j_{n}m_{n}}+(-1)^{j_{n}-m_{n}}V_{j_{n}}\alpha_{j_{n}-m_{n},}\nonumber
\end{eqnarray}
where $U_{j_{n}} (U_{j_{p}})$, $V_{j_{n}} (V_{j_{p}})$,
$a^{+}_{j_{n}m_{n}}(a_{j_{p}m_{p}})$ and
$\alpha^{+}_{j_{n}m_{n}}(\alpha_{j_{p}m_{p}})$ are the standard BCS
occupation amplitudes, the nucleon creation (annihilation) and the
quasi-particle creation (annihilation) operators, respectively. We
considered a system of nucleons in a deformed mean field with
pairing forces. The single quasi-particle (sqp) Hamiltonian of the
system can be defined using
\begin{eqnarray}
\hat{H}_{sqp}=\sum_{j_{\tau}m_{\tau}}[\varepsilon_{j_{n}}\alpha^{+}_{j_{n}m_{n}}\alpha_{j_{n}m_{p}}+\varepsilon_{j_{p}}\alpha^{+}_{j_{p}m_{p}}\alpha_{j_{p}m_{p}}],~~~~\tau=n,p\nonumber
\end{eqnarray}
where $\varepsilon_{j_{\tau}}$ are the energies of neutron(proton)
quasi-particle
\begin{eqnarray}
\varepsilon_{j_{\tau}}=\sqrt{(E_{j_{\tau}}-\lambda)^{2}+\Delta^{2}}\nonumber.
\end{eqnarray}
Here $E_{j_{\tau}}$, $\lambda$ and $\Delta$ are the single particle
neutron(proton) energy, the chemical potential and pairing energies,
respectively. Charge-exchange spin-spin correlations are added to
the model Hamiltonian in the following form
\begin{eqnarray}
\hat{V}_{\beta}=2\chi_{\beta}\sum_{\beta}\beta^{+}_{\mu}\beta^{-}_{\mu},~~~~~~\mu=0,\pm 1\nonumber
\end{eqnarray}
where $\beta^{+}_{\mu}(\beta^{-}_{\mu})$ is the positron(electron)
decay operator
\begin{eqnarray}
\beta^{+}_{\mu}=\sum_{n\rho}\sum_{p\rho^{'}}<n\rho|\sigma_{\mu}+(-1)^{\mu}\sigma_{-\mu}|p\rho^{'}>a^{+}_{n\rho}a_{p\rho^{'}},~~~~~~\beta^{-}_{\mu}=({\beta^{+}_{\mu}})^{\dag}\nonumber
\end{eqnarray}
and $\sigma_{\mu}$ is and the spherical component of the Pauli
operator. The main formulae are given here. Details mathematical
formalism are available in \cite{Gab70,Ayg02,Sel03,Sel04}.

The total pn-QRPA Hamiltonian for deformed nuclei is described as
\begin{eqnarray}
\hat{H}_{DSM}=\hat{H}_{sqp}+\hat{V}_{\beta}.
\end{eqnarray}
The total Hamiltonian in Pyatov's method is given by
\begin{eqnarray}
\hat{H}_{SPM}=\hat{H}_{av}+\hat{V}_{\beta}+\hat{h}_{0},
\end{eqnarray}
where $\hat{H}_{av}$ is the single quasi-particle Hamiltonian in a
spherical symmetric average field with pairing forces. The third
term comes from the restoration of broken commutation relation
between the nuclear Hamiltonian and the GT operator. The schematic
method Hamiltonian for GT excitations in the neighbor odd-odd nuclei
is given by
\begin{eqnarray}
\hat{H}_{SSM}=\hat{H}_{av}+\hat{V}_{\beta}.
\end{eqnarray}
Details of solution of allowed GT formalism can be seen in
\cite{Cak10a,Cak12}.

The ft values for the allowed GT $\beta$ transitions are finally
calculated using
\begin{eqnarray}
ft=\frac{D}{(\frac{g_{A}}{g_{V}})^{2}4\pi B^{GT}(I_{i}\rightarrow I_{f},\beta^{-})},\nonumber
\end{eqnarray}
where the reduced matrix elements of GT transitions are given by
\begin{eqnarray}
B^{GT}(I_{i}\rightarrow I_{f},\beta^{-})=\sum_{\mu}|\langle1^{+}_{i},\mu|G_{\mu}^{-}|0^{+}\rangle|^{2}\nonumber.
\end{eqnarray}

The model Hamiltonian which generates the spin-isospin dependent
vibration modes with $\lambda^{\pi}=0^{-},1^{-},2^{-}$ in odd-odd
nuclei in quasi boson approximation is given as
\begin{eqnarray}
\hat{H}=\hat{H}_{sqp}+\hat{h}_{ph}.
\end{eqnarray}
The single quasi-particle Hamiltonian of the system is given by
\begin{eqnarray}
\hat{H}_{sqp}=\sum_{j_{\tau}}\varepsilon_{j_{\tau}}\alpha^{\dag}_{j_{\tau}m_{\tau}}\alpha_{j_{\tau}m_{\tau}}~~(\tau=p,n),\nonumber
\end{eqnarray}
where $\varepsilon_{j_{\tau}}$ and $\alpha^{+}_{j_{\tau}m_{\tau}}(\alpha_{j_{\tau}m_{\tau}})$ are the single quasi-particle energy of the nucleons with angular momentum ${j_{\tau}}$ and the quasi-particle creation (annihilation) operators, respectively.

The $\hat{h}_{ph}$ is the spin-isospin effective interaction
Hamiltonian which generates $0^{-}$,$1^{-}$,$2^{-}$ vibration modes
in particle-hole channel and given as
\begin{eqnarray}
\hat{h}_{ph}=\frac{2\chi_{ph}}{g_{A}^{2}}\sum_{j_{p}j_{n}j_{p'}j_{n'}\mu}[b_{j_{p}j_{n}}A^{+}_{j_{p}j_{n}}(\lambda\mu)+(-1)^{\lambda-\mu}\bar{b}_{j_{p}j_{n}}A_{j_{p}j_{n}}(\lambda-\mu)]\nonumber
\end{eqnarray}
\begin{eqnarray}
[b_{j_{p'}j_{n'}}A_{j_{p'}j_{n'}}(\lambda\mu)+(-1)^{\lambda-\mu}\bar{b}_{j_{p'}j_{n'}}A^{+}_{j_{p'}j_{n'}}(\lambda-\mu)]\nonumber
\end{eqnarray}
where $\chi_{ph}$ is particle-hole effective interaction constant.

The quasi-boson creation $A^{+}_{j_{p}j_{n}}(\lambda\mu)$ and
annihilation $A_{j_{p}j_{n}}(\lambda\mu)$ operators are given as
\begin{eqnarray}
A^{+}_{j_{p}j_{n}}(\lambda\mu)=\sqrt{\frac{2\lambda+1}{2j_{p}+1}}\sum_{m_{n}m_{p}}(-1)^{j_{n}-m_{n}}\langle j_{n}m_{n}\lambda\mu | j_{p}m_{p}\rangle\alpha^{+}_{j_{p}m_{p}}\alpha^{+}_{j_{n}-m_{n}},~~~~~A_{j_{p}j_{n}}(\lambda\mu)=\{A^{+}_{j_{p}j_{n}}(\lambda\mu)\}^{\dag}.\nonumber
\end{eqnarray}

The $b_{j_{p}j_{n}}$, $\bar{b}_{j_{p}j_{n}}$ are the reduced matrix
elements of the non-relativistic multipole operators for rank 0, 1
and 2 \cite{Boh69} and given by
\begin{eqnarray}
b_{j_{p}j_{n}}=<j_{p}(l_{p}s_{p})\|r_{k}[Y_{1}\sigma_{k}]_{0}\|j_{n}(l_{n}s_{n})>V_{j_{n}}U_{j_{p}},~~~~~\bar{b}_{j_{p}j_{n}}=<j_{p}(l_{p}s_{p})\|r_{k}[Y_{1}\sigma_{k}]_{0}\|j_{n}(l_{n}s_{n})>U_{j_{n}}V_{j_{p}},\nonumber
\end{eqnarray}
\begin{eqnarray}
b_{j_{p}j_{n}}=<j_{p}(l_{p}s_{p})\|r_{k}[Y_{1}\sigma_{k}]_{1}\|j_{n}(l_{n}s_{n})>V_{j_{n}}U_{j_{p}},~~~~~\bar{b}_{j_{p}j_{n}}=<j_{p}(l_{p}s_{p})\|r_{k}[Y_{1}\sigma_{k}]_{1}\|j_{n}(l_{n}s_{n})>U_{j_{n}}V_{j_{p}},\nonumber
\end{eqnarray}
\begin{eqnarray}
b_{j_{p}j_{n}}=<j_{p}(l_{p}s_{p})\|r_{k}[Y_{1}\sigma_{k}]_{2}\|j_{n}(l_{n}s_{n})>V_{j_{n}}U_{j_{p}},~~~~~\bar{b}_{j_{p}j_{n}}=<j_{p}(l_{p}s_{p})\|r_{k}[Y_{1}\sigma_{k}]_{2}\|j_{n}(l_{n}s_{n})>U_{j_{n}}V_{j_{p}},\nonumber
\end{eqnarray}
where $U_{j_{\tau}}$ and $V_{j_{\tau}}$ are the standard BCS
occupation amplitudes. The calculation of the transition
probabilities for rank 0 and rank 1 have been performed using $\xi$
approximation (see \cite{Boh69} for a detailed information about the
$\xi$ approximation).

Calculation of rank 0 FF transitions was done within the
pn-QRPA(WS-SSM) formalism. Details of this calculation can be seen
from \cite{Cak10}. The first forbidden transitions are dictated by
the matrix elements of moments \cite{Boh69}. The relativistic and
the non-relativistic matrix elements, respectively, for
$\lambda^{\pi}=0^{-}$ are given by
\begin{eqnarray}
M^{\mp}(\rho_{A},\lambda=0)=\frac{g_{A}}{\sqrt{4\pi}c}\sum_{k}t_{\mp}(k)(\vec{\sigma}_{k}\cdot\vec\vartheta_{k}),\nonumber
\end{eqnarray}
\begin{eqnarray}
M^{\mp}(j_{A},\kappa=1,\lambda=0)=g_{A}\sum_{k}t_{\mp}(k)r_{k}\{Y_{1}(r_{k})\sigma_{k}\}_{0}.\nonumber
\end{eqnarray}
The relativistic and the non-relativistic matrix elements, respectively, for $\lambda^{\pi}=1^{-}$ are given by
\begin{eqnarray}
M^{\mp}(j_{v},\kappa=0,\lambda=1,\mu)=\frac{g_{v}}{\sqrt{4\pi}c}\sum_{k}t_{\mp}(k)r_{k}(\vec\vartheta_{k})_{1\mu},\nonumber
\end{eqnarray}
\begin{eqnarray}
M^{\mp}(\rho_{v},\lambda=1,\mu)=g_{v}\sum_{k}t_{\mp}(k)r_{k}Y_{1\mu}(r_{k}),\nonumber
\end{eqnarray}
\begin{eqnarray}
M^{\mp}(j_{v},\kappa=1,\lambda=1,\mu)=g_{A}\sum_{k}t_{\mp}(k)r_{k}\{Y_{1}(r_{k})\sigma_{k}\}_{1\mu}.\nonumber
\end{eqnarray}
Finally, the non-relativistic matrix element for $\lambda^{\pi}=2^{-}$ is given as
\begin{eqnarray}
M^{\mp}(j_{A},\kappa=1,\lambda=2,\mu)=g_{A}\sum_{k}t_{\mp}(k)r_{k}\{Y_{1}(r_{k})\sigma_{k}\}_{2\mu}.\nonumber
\end{eqnarray}

The transitions probabilities $B(\lambda^{\pi}=0^{-},1^{-},2^{-}; \beta^{\mp})$ are given by \cite{Boh69}
\begin{eqnarray}
B(\lambda^{\pi}=0^{-},\beta^{\mp})=|<0^{-}_{i}\|M_{\beta^{\mp}}^{0}\|0^{+}>|^{2},\nonumber
\end{eqnarray}
where
\begin{eqnarray}
M_{\beta^{\mp}}^{0}=\pm
M^{\mp}(\rho_{A},\lambda=0)-i\frac{m_{e}c}{\hbar}\xi
M^{\mp}(j_{A},\kappa=1,\lambda=0).
\end{eqnarray}
\begin{eqnarray}
B(\lambda^{\pi}=1^{-},\beta^{\mp})=|<1^{-}_{i}\|M_{\beta^{\mp}}^{1}\|0^{+}>|^{2},\nonumber
\end{eqnarray}
where
\begin{eqnarray}
M_{\beta^{\mp}}^{1}=M^{\mp}(j_{v},\kappa=0,\lambda=1,\mu)\pm
i\frac{m_{e}c}{\sqrt{3}\hbar}
M^{\mp}(\rho_{v},\lambda=1,\mu)+i\sqrt{\frac{2}{3}}\frac{m_{e}c}{\hbar}\xi
M^{\mp}(j_{A},\kappa=1,\lambda=1,\mu).
\end{eqnarray}
\begin{eqnarray}
B(\lambda^{\pi}=2^{-},\beta^{\mp})=|<2^{-}_{i}\|M_{\beta^{\mp}}^{2}\|0^{+}>|^{2},\nonumber
\end{eqnarray}
where
\begin{eqnarray}
M_{\beta^{\mp}}^{2}= M^{\mp}(j_{A},\kappa=1,\lambda=2,\mu).
\end{eqnarray}
In Eq.(5) and Eq.(6), the upper and lower signs refer to $\beta^{-}$ and $\beta^{+}$ decays, respectively.

The ft values are given by the following expression:
\begin{eqnarray}
(ft)_{\beta^{\mp}}=\frac{D}{(g_{A}/g_{V})^{2}4\pi B(I_{i}\longrightarrow I_{f}, \beta^{\mp})}\nonumber
\end{eqnarray}
where
\begin{eqnarray}
D=\frac{2\pi^{3}\hbar^{2}ln2}{g_{v}^{2}m_{e}^{5}c^{4}}=6250~sec,~~~\frac{g_{A}}{g_{v}}=-1.254.\nonumber
\end{eqnarray}
Transitions with $\lambda=n+1$ are referred to as unique first forbidden transitions \cite{Boh69}, and the $ft$ values are expressed as
\begin{eqnarray}
(ft)_{\beta^{\mp}}=\frac{D}{(g_{A}/g_{V})^{2}4\pi B(I_{i}\longrightarrow I_{f}, \beta^{\mp})}~\frac{(2n+1)!!}{[(n+1)!]^{2}n!}.\nonumber
\end{eqnarray}

\subsection{The pn-QRPA(N) model}
In the pn-QRPA(N) formalism \cite{Mut92}, proton-neutron residual
interactions occur as particle-hole (characterized by interaction
constant $\chi$) and particle-particle (characterized by interaction
constant $\kappa$) interactions. The particle-particle interaction
was usually neglected in previous $\beta^{-}$-decay calculations
\cite{Kru84,Moe90,Ben88,Sta89,Sta90}. However it was later found to
be important, specially for the calculation of $\beta^{+}$-decay
\cite{Sta90,Suh88,Hir91,Hir93}. The incorporation of
particle-particle force leads to a redistribution of the calculated
$\beta$ strength, which is commonly shifted toward lower excitation
energies \cite{Hir93}. We use a schematic separable interaction. The
advantage of using these separable GT forces is that the QRPA matrix
equation reduces to an algebraic equation of fourth order, which is
much easier to solve as compared to full diagonalization of the
non-Hermitian matrix of large dimensionality \cite{Hom96,Mut92}.

Essentially we first constructed a quasiparticle basis (defined by a
Bogoliubov transformation) with a pairing interaction, and then
solved the RPA equation with a schematic separable GT residual
interaction. As a starting point, single-particle energies and wave
functions are calculated in the Nilsson model, which takes into
account nuclear deformation. The transformation from the spherical
basis to the axial-symmetric deformed basis is governed by
\begin{eqnarray}\label{sd}
d^{\dagger}_{m\alpha}=\Sigma_{j}D^{m\alpha}_{j}s^{\dagger}_{jm},
\end{eqnarray}
where $d^{+}$ and $s^{+}$ are particle creation operators in the
deformed and spherical basis, respectively, and the matrices
$D^{m\alpha}_{j}$ are determined by diagonalization of the Nilsson
Hamiltonian. The BCS calculation was performed in the deformed
Nilsson basis for neutrons and protons separately. We employed a
constant pairing force and introduced a quasiparticle basis via
\begin{eqnarray}\label{qbas}
a^{\dagger}_{m\alpha}=u_{m\alpha}d^{\dagger}_{m\alpha}-v_{m\alpha}d_{\bar{m}\alpha}
\\ \nonumber
a^{\dagger}_{\bar{m}\alpha}=u_{m\alpha}d^{\dagger}_{\bar{m}\alpha}+v_{m\alpha}d_{m\alpha}
\end{eqnarray}
where $\bar{m}$ is the time reversed state of $m$ and
$a^{\dagger}/a$ are the quasiparticle creation/annihilation
operators which enter the RPA equation. The occupation amplitudes
$u$ and $v$ satisfy the condition $u^{2} + v^{2} = 1$ and are
determined from the BCS equaitons. The formalism for solving the RPA
equation and calculation of allowed $\beta$-decay rates in stellar
matter using the pn-QRPA(N) model can be seen in detail from
\cite{Hir93}. Below we describe briefly the necessary formalism to
calculate the unique FF (referred to as U1F) $\beta$-decay rates.

For the calculation of the U1F $\beta$-decay rates, nuclear matrix
elements of the separable forces which appear in RPA equation are
given by

\begin{equation}\label{vph}
V^{ph}_{pn,p^{\prime}n^{\prime}} = +2\chi
f_{pn}(\mu)f_{p^{\prime}n^{\prime}}(\mu),
\end{equation}

\begin{equation}\label{vpp}
V^{pp}_{pn,p^{\prime}n^{\prime}} = -2\kappa
f_{pn}(\mu)f_{p^{\prime}n^{\prime}}(\mu),
\end{equation}

where
\begin{equation}\label{fpn}
f_{pn}(\mu)=<j_{p}m_{p}|t_{-}r[\sigma Y_{1}]_{2\mu}|j_{n}m_{n}>,
\end{equation}
is a single-particle U1F transition amplitude (the symbols have
their normal meaning). Note that $\mu$ takes the values
$\mu=0,\pm1$, and $\pm2$ (for allowed decay rates $\mu$ only takes
the values $0$ and $\pm1$), and the proton and neutron states have
opposite parities \cite{Hom96}.

Choice of particle-particle and particle-hole interaction strength
needs special mention. Range of values for $\chi$ is roughly from
0.001 to 0.8 and $\kappa$ roughly from 0. to 0.15 in earlier
calculations of pn-QRPA(N) where a locally best value of $\chi$ and
$\kappa$ was found for every isotopic chain \cite{Sta90,Hir93}. Of
course one has to ensure that for large values of $\kappa$, the
model does not "collapse" (i.e. lowest eigenvalue does not become
complex). In initial works, we sought a Z-dependent value of $\chi$
and $\kappa$ for Fe-isotopes \cite{Nab10,Nab11,Nab11a} and
Ni-isotopes \cite{Nab12,Nab13,Nab13a}. However, in literature,
$A$-dependent values of $\chi$ and $\kappa$ are also frequently
cited for RPA methods (e.g. \cite{Civ86,Hom96,Moe03}). Our recent
findings show that a mass-dependent $\chi$ and $\kappa$ formula
better reproduces the experimental half-lives specially for cases
where contributions from FF decays are also taken into account.
Accordingly in this work we searched for mass-dependent $\chi$ and
$\kappa$ values for even-even neutron-rich isotopes of nickel and
found $\chi$= $4.2/A$ MeV for allowed and $56.16/A$ MeV fm$^{-2}$
for U1F transitions.  The other interaction constant $\kappa$ was
taken to be zero. These values of $\chi$ and $\kappa$ best
reproduced the measured half-lives. The same values of $\chi$ and
$\kappa$ were also used in another recent calculation of FF
$\beta$-decay rates of Zn and Ge isotopes \cite{Nab15}.

Deformation of the nuclei was calculated using
\begin{equation}\label{del}
\delta = \frac{125(Q_{2})}{1.44 (Z) (A)^{2/3}},
\end{equation}
where $Z$ and $A$ are the atomic and mass numbers, respectively and
$Q_{2}$ is the electric quadrupole moment taken from Ref.
\cite{Moe81}. Q-values were taken from the mass compilation of Audi
et al. \cite{Aud12}.

We are currently working on calculation of rank 0 FF transition
phase space factors at finite temperatures. This would be treated as
a future assignment and currently we are only able to calculate
phase factor for rank 2 forbidden (U1F) transitions under stellar
conditions. The U1F stellar $\beta$-decay rates from the
$\mathit{i}$th state of the parent to the $\mathit{j}$th state of
the daughter nucleus is given by

\begin{equation}\label{lij}
\lambda_{ij}^{\beta} =
\frac{m_{e}^{5}c^{4}}{2\pi^{3}\hbar^{7}}\sum_{\Delta
J^{\pi}}g^{2}f_{ij}(\Delta J^{\pi})B_{ij}(\Delta J^{\pi}),
\end{equation}
where $f_{ij}(\Delta J^{\pi})$ and $ B_{ij}(\Delta J^{\pi})$ are the
integrated Fermi function and the reduced transition probability for
$\beta$-decay, respectively given as

\begin{eqnarray}\label{e}
B_{ij}(\Delta
J^{\pi})=\frac{1}{12}z^{2}(w_{m}^{2}-1)-\frac{1}{6}z^{2}w_{m}w+\frac{1}{6}z^{2}w^{2},
\end{eqnarray}
where $z$ is
\begin{eqnarray}\label{efg}
z=2g_{A}\frac{\langle f|\vert\sum_{k}r_{k}[\textbf{C}^{k}_{1}\times
\boldsymbol{\sigma}]^{2}{\textbf{t}}^{k}_{-}\vert|i\rangle}{\sqrt{2J_{i}+1}},
\end{eqnarray}
where
\begin{eqnarray}\label{gkl}
\textbf{C}_{lm}=\sqrt{\frac{4\pi}{2l+1}}\textbf{Y}_{lm},
\end{eqnarray}
with $\textbf{Y}_{lm}$ the spherical harmonics. The phase space
integral $f_{ij}$ for U1F transitions can be obtained as
\begin{eqnarray}\label{f}
f_{ij} = \int_{1}^{w_{m}} w \sqrt{w^{2}-1}
(w_{m}-w)^{2}[(w_{m}-w)^{2}F_{1}(Z,w) \nonumber\\
+ (w^{2}-1)F_{2}(Z,w)] (1-G_{-}) dw,
\end{eqnarray}
where $w$ is the total kinetic energy of the electron including its
rest mass and $w_{m}$ is the total $\beta$-decay energy ($ w_{m} =
m_{p}-m_{d}+E_{i}-E_{j}$, where $m_{p}$ and $E_{i}$ are mass and
excitation energies of the parent nucleus, and $m_{d}$ and $E_{j}$
of the daughter nucleus, respectively). $G_{-}$ are the electron
distribution functions. Assuming that the electrons are not in a
bound state, these are the Fermi-Dirac distribution functions,
\begin{equation}\label{Gm}
G_{-} = [exp
(\frac{E-E_{f}}{kT})+1]^{-1}.
\end{equation}
Here $E=(w-1)$ is the kinetic energy of the electrons, $E_{f}$ is
the Fermi energy of the electrons, $T$ is the temperature, and $k$
is the Boltzmann constant.

The Fermi functions, $F_{1}(\pm Z,w)$ and $F_{2}(\pm Z,w)$ appearing
in Eq.~(\ref{f}) were calculated according to the procedure adopted
by \cite{Gov71}.

The number density of electrons associated with protons and nuclei
is $\rho Y_{e} N_{A}$, where $\rho$ is the baryon density, $Y_{e}$
is the ratio of electron number to the baryon number, and $N_{A}$ is
the Avogadro's number.
\begin{equation}\label{ye}
\rho Y_{e} = \frac{1}{\pi^{2}N_{A}}(\frac {m_{e}c}{\hbar})^{3}
\int_{0}^{\infty} (G_{-}-G_{+}) p^{2}dp,
\end{equation}
where $p=(w^{2}-1)^{1/2}$ is the electron or positron momentum, and
Eq.~(\ref{ye}) has the units of \textit{moles $cm^{-3}$}. $G_{+}$
are the positron distribution functions given by
\begin{equation}\label{gp}
G_{+} =\left[\exp
\left(\frac{E+2+E_{f} }{kT}\right)+1\right]^{-1}.
\end{equation}
Eq.~(\ref{ye}) is used for an iterative calculation of Fermi
energies for selected values of $\rho Y_{e}$ and $T$.

There is a finite probability of occupation of parent excited states
in the stellar environment as a result of the high temperature in
the interior of massive stars. Weak decay rates then also have a
finite contribution from these excited states. The occupation
probability of a state $i$ is calculated on the assumption of
thermal equilibrium,

\begin{equation}\label{pi}
P_{i} = \frac {exp(-E_{i}/kT)}{\sum_{i=1}exp(-E_{i}/kT)},
\end{equation}
where $E_{i}$ is the excitation energy of the state $i$,
respectively. The rate per unit time per nucleus for stellar
$\beta$-decay process is finally given by
\begin{equation}\label{lb}
\lambda^{\beta} = \sum_{ij}P_{i}
\lambda_{ij}^{\beta}.
\end{equation}
The summation over all initial and final states are carried out
until satisfactory convergence in the rate calculations is achieved.
We note that due to the availability of a huge model space (up to 7
major oscillator shells) convergence is easily achieved in our rate
calculations for excitation energies well in excess of 10 MeV (for
both parent and daughter states).

It is assumed in our calculation that all daughter excited states,
with energy greater than the separation energy of neutrons ($S_{n}$)
decay by emission of neutrons. The neutron energy rate from the
daughter nucleus is calculated using
\begin{equation}\label{ln}
\lambda^{n} = \sum_{ij}P_{i}\lambda_{ij}(E_{j}-S_{n}),
\end{equation}
for all $E_{j} > S_{n}$.

The probability of $\beta$-delayed neutron emission is calculated by
\begin{equation}\label{pn}
P^{n} =
\frac{\sum_{ij\prime}P_{i}\lambda_{ij\prime}}{\sum_{ij}P_{i}\lambda_{ij}},
\end{equation}
where $j\prime$ are states in the daughter nucleus for which
$E_{j\prime} > S_{n}$. In Eq.~(\ref{ln}) and Eq.~(\ref{pn}),
$\lambda_{ij(\prime)}$ is the sum of the positron capture and
electron decay rates, for the transition $i$ $\rightarrow$
$j(j\prime)$.

\section{Results and comparison}

The calculated GT strength  transitions in our pn-QRPA(WS-SSM),
pn-QRPA(WS-SPM) and pn-QRPA(WS-DSM) models are shown in
Table~\ref{ta1} for chosen nickel isotopes. The inclusion of
deformation lifts the degeneracy of energy levels presented in the
spherical models.  Only particle-hole interaction strength  was
considered for both allowed GT and FF calculations within the
pn-QRPA(WS) formalism. A quenching factor of 0.6 was applied for all
pn-QRPA(WS) and pn-QRPA(N) calculations. The pairing correlation
constants were taken as $C_{n}=C_{p}=12 /\sqrt{A}$. The excitation
energies shown in Table~\ref{ta1} are the ones obtained over the
ground states of daughter nuclei.  The strength parameters of the
effective interaction are $\chi_{\beta}=5.2A^{0.7} MeV$
\cite{Hom96}.

The calculated FF charge-changing transition strengths using the
pn-QRPA(WS-SSM) model are shown in Table~\ref{ta2}.  The strength
parameters of the effective interaction are $\chi_{\beta}=30A^{-5/3}
MeVfm^{-2}$, $\chi_{\beta}=55A^{-5/3} MeVfm^{-2}$ and
$\chi_{\beta}=99A^{-5/3} MeVfm^{-2}$ for rank0, rank1 and rank2,
respectively.

\begin{figure}[p]
\begin{center}
\includegraphics[width=1.0\textwidth]{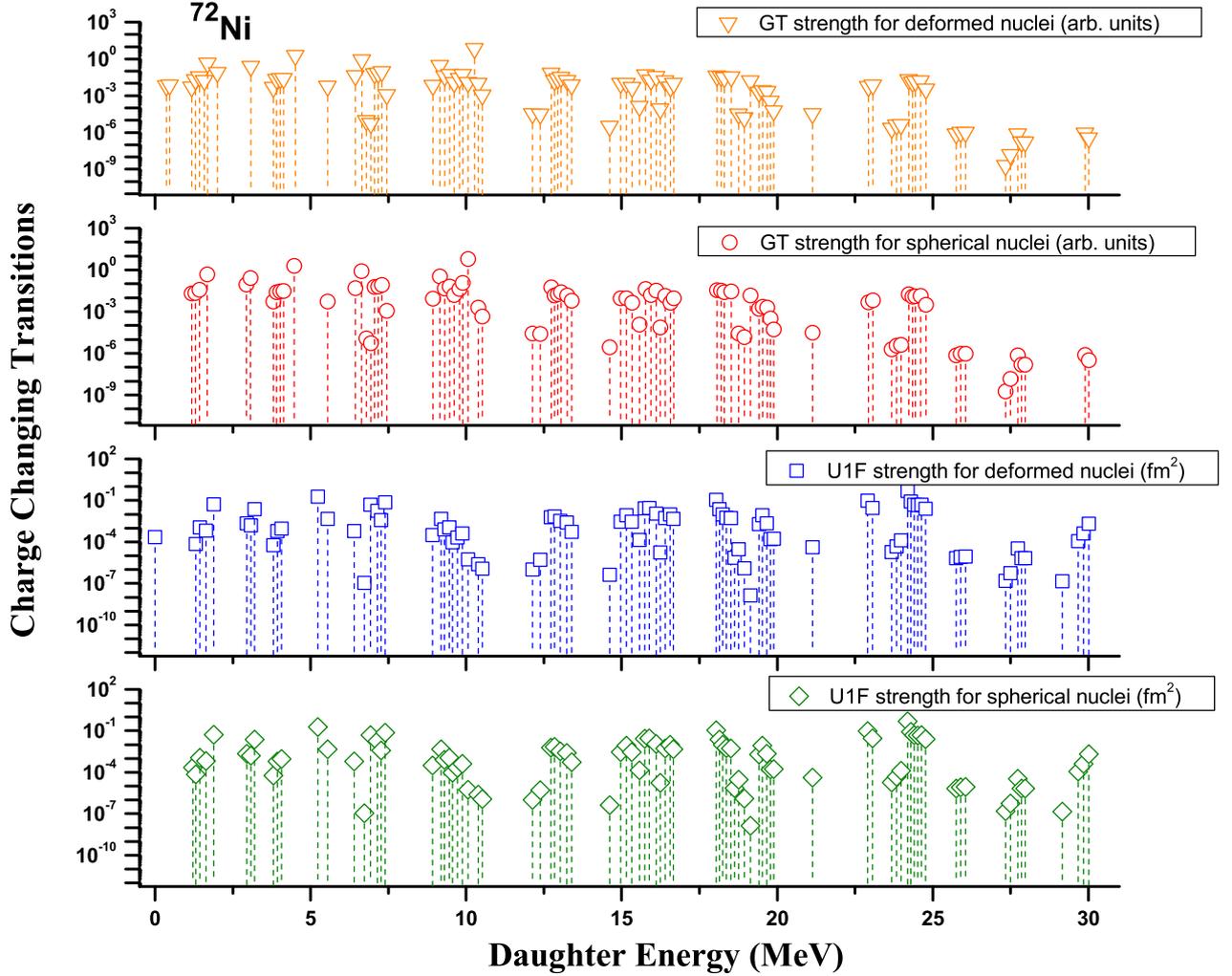}
\caption{\scriptsize Allowed (GT) and  unique first-forbidden (U1F)
charge-changing transition strengths calculated in the pn-QRPA(N)
model for deformed and spherical cases of $^{72}$Ni.}
\label{figure1}
\end{center}
\end{figure}

\begin{figure}[p]
\begin{center}
\includegraphics[width=1.0\textwidth]{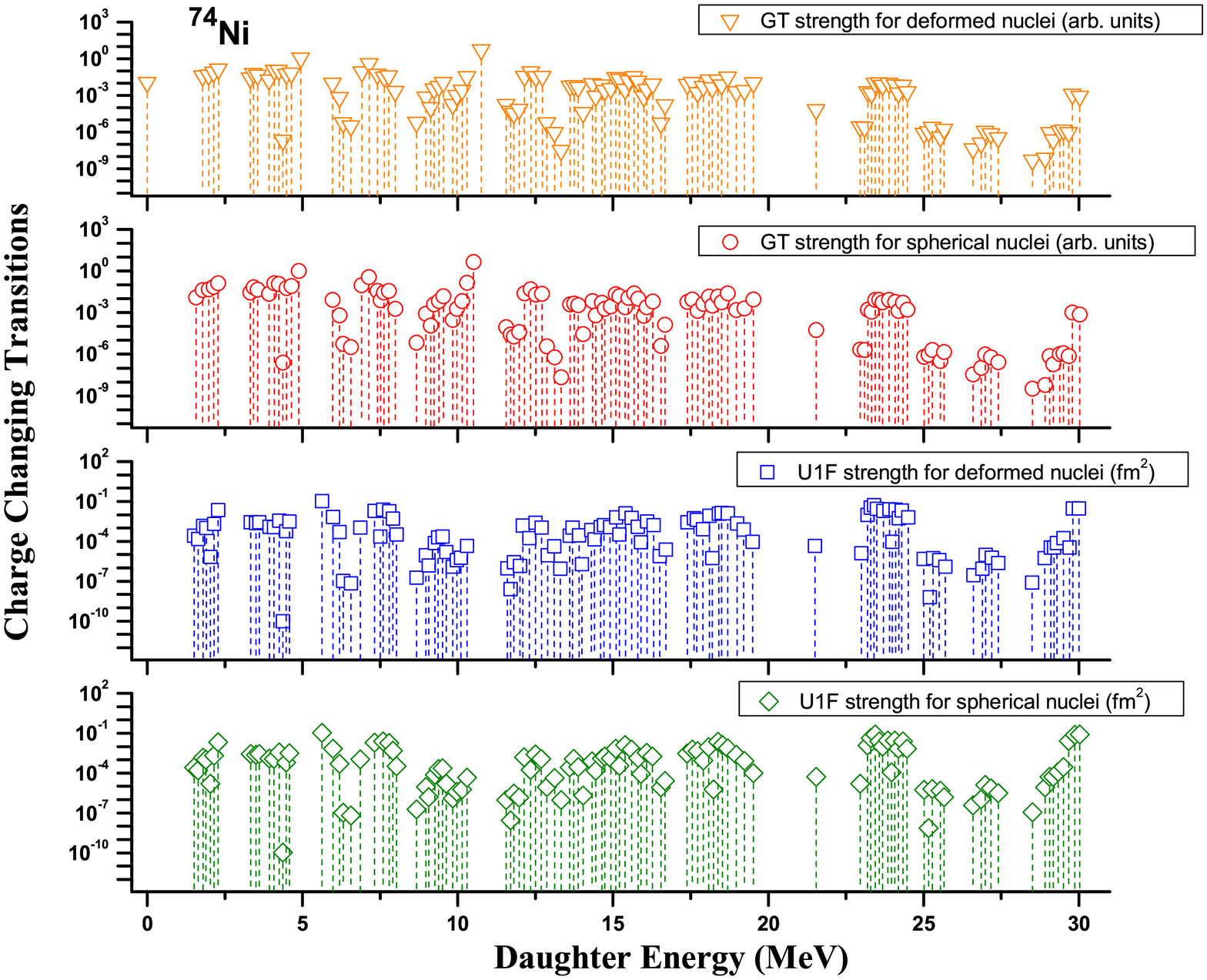}
\caption{\scriptsize Same as Fig.~\ref{figure1} but for $^{74}$Ni.}
\label{figure2}
\end{center}
\end{figure}

\begin{figure}[p]
\begin{center}
\includegraphics[width=1.0\textwidth]{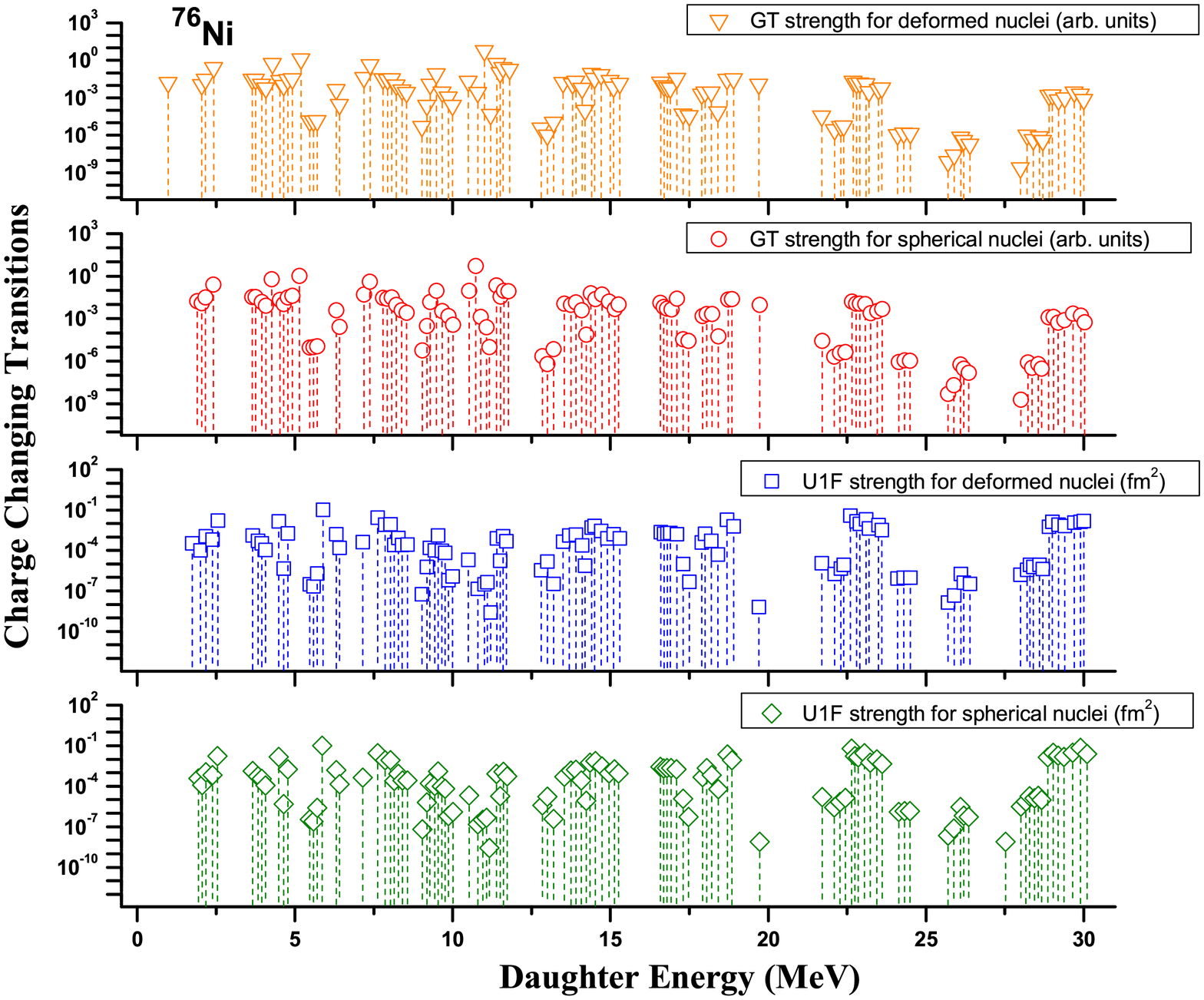}
\caption{\scriptsize Same as Fig.~\ref{figure1} but for $^{76}$Ni.}
\label{figure3}
\end{center}
\end{figure}

\begin{figure}[p]
\begin{center}
\includegraphics[width=1.0\textwidth]{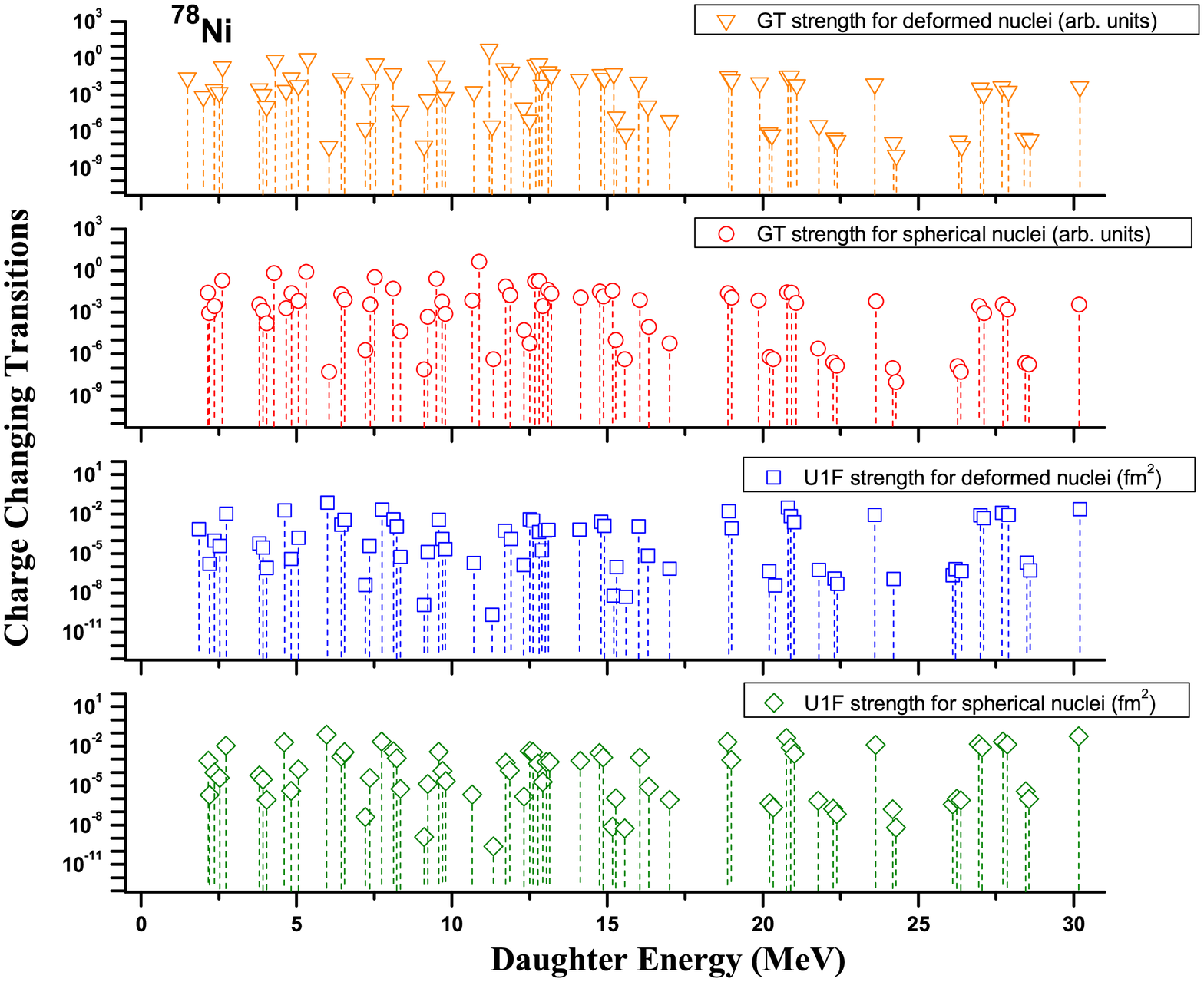}
\caption{\scriptsize Same as Fig.~\ref{figure1} but for $^{78}$Ni.}
\label{figure4}
\end{center}
\end{figure}

The deformation parameters for the even-even isotopes of Ni used in
both pn-QRPA(WS) and pn-QRPA(N) models are shown in Table~\ref{ta3}.
In order to study the effect of deformation in Nilsson calculation,
we performed two sets of calculation within the pn-QRPA(N) model. In
the first case the deformation parameter was taken as zero and in
the second case the deformation was taken from Table~\ref{ta3}. The
resulting charge-changing transitions both for allowed and U1F
transitions are shown in Figs.~\ref{figure1}-\ref{figure4} for
$^{72}$Ni, $^{74}$Ni, $^{76}$Ni and $^{78}$Ni, respectively. The
small value of deformation does not appreciably change the
calculated strength distributions except for a few states in the
low-lying energy region. The calculated allowed GT and U1F
$\beta$-decay rates for the spherical and deformed cases were almost
the same. Only at high stellar temperatures (T$_{9} \geq$ 10) did
the decay rate for the deformed case increase by around 10$\%$ as
against those cases where we treated the nuclei as spherical.

\begin{figure}[p]
\begin{center}
\includegraphics[width=1.0\textwidth]{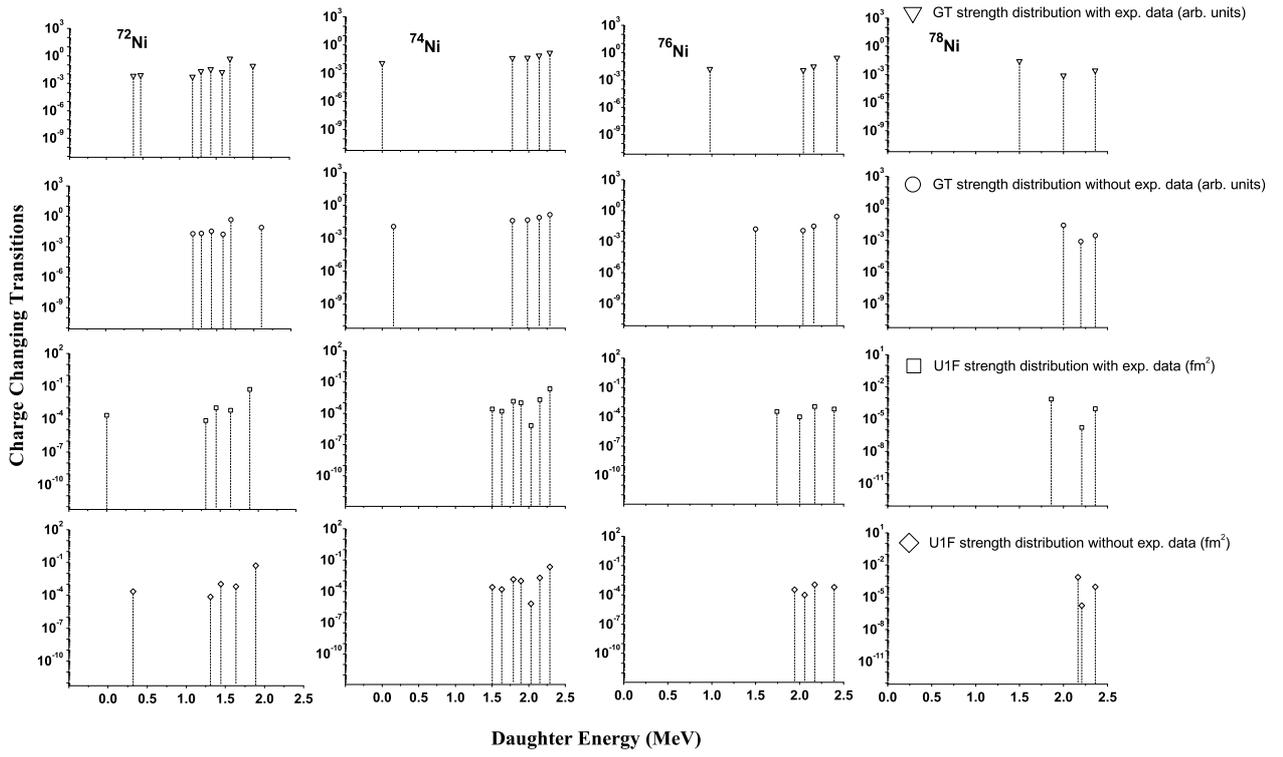}
\caption{\scriptsize Allowed (GT) and  unique first-forbidden (U1F)
charge-changing transition strengths calculated in the pn-QRPA(N)
model before and after insertion of experimental data.}
\label{figure5}
\end{center}
\end{figure}

Insertion of experimental data in the pn-QRPA(N) model deserves
special mention. If the original calculated charge-changing strength
distribution differs considerably from those after insertion of
experimental data then it undermines the predictive power of the
pn-QRPA(N) model. Fig.~\ref{figure5} shows the calculated strength
distributions before and after insertion of experimental data both
for allowed GT and U1F transitions. In case of allowed GT strength
distribution for $^{72}$Ni, the first calculated state is fragmented
into three states (two of which are placed at experimentally
measured levels). In other cases there is  displacement of first few
energy levels within 500 keV (which is roughly the uncertainty in
calculation of energy eigenvalues in the pn-QRPA(N) model).  Further
there are no experimental insertions beyond 2.5 MeV in daughter
energy in all nickel isotopes. The predictive power of the
pn-QRPA(N) model gets better for shorter half-lives, that is, with
increasing distance from line of stability \cite{Hir93}.

The allowed GT $\beta$-decay half-lives of nickel isotopes
calculated within the pn-QRPA formalism are shown in
Table~\ref{ta6}. The calculated half-lives are also compared with
experimental data. The recent atomic mass evaluation data of Ref.
\cite{Aud12} have been used for experimental half-lives values. The
calculated half-lives of M\"{o}ller et al from \cite{Moe97} are also
shown in comparison which uses the deformation of nucleus and
folded-Yukawa single-particle potential. It may be concluded that
QRPA calculation of M\"{o}ller et al. improves as the nucleus
becomes more neutron-rich. It can be seen that the pn-QRPA(N) and
pn-QRPA(WS-DSM) models calculate half-lives in better agreement with
the measured half-lives.

The FF contributions to the total calculated half-lives are shown in
Table~\ref{ta7}. Here we present the GT+U1F calculation of
half-lives in the pn-QRPA(N) model and the GT+rank0+rank1+rank2
half-lives calculation using the pn-QRPA(WS-SSM) model. We are
currently working on rank1 and rank 2 calculation of half-lives in
the pn-QRPA(N) model. Calculation of FF contributions in
pn-QRPA(WS-SPM) and pn-QRPA(WS-DSM) models would also be taken as a
future assignment. Table~\ref{ta7} shows that the calculated
half-lives get appreciably smaller and in better agreement with
measured half-lives when the U1F contribution is added in the
pn-QRPA(N) model. Likewise, but to a smaller extent, the half-lives
get smaller once the FF contributions are added in the
pn-QRPA(WS-SSM) model.  Contribution of U1F rates to total
$\beta$-decay half-lives in the pn-QRPA(N) model is 19.7$\%$,
24.0$\%$, 18.5$\%$ and 17.5$\%$ for $^{72}$Ni, $^{74}$Ni, $^{76}$Ni
and $^{78}$Ni, respectively.  Calculated half-lives of \cite{Bor05}
using the DF3 + CQRPA model (which includes the FF contribution) can
also be seen in Table~\ref{ta7}. The DF3 + CQRPA results get in
better agreement with experimental data as $N$ (neutron number)
increases. It is concluded that pn-QRPA(N) emerges as the best model
and has overall excellent agreement with experimentally determined
half-lives of Ni isotopes. It is also expected to give reliable
results for nuclei close to neutron-drip line for which no
experimental data is available.

\begin{figure}[p]
\begin{center}
\includegraphics[width=1.0\textwidth]{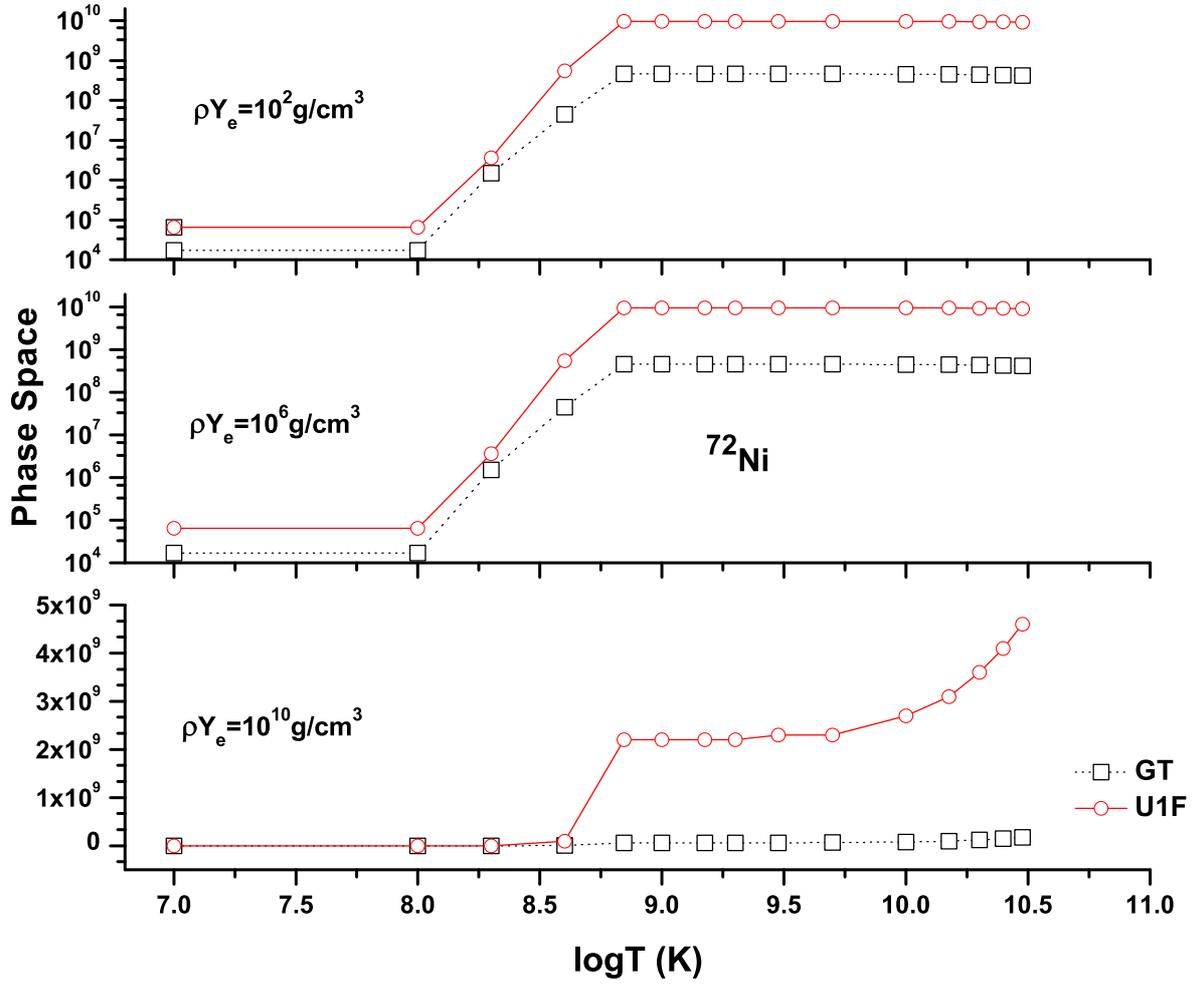}
\caption{\scriptsize Comparison of calculated phase space for
allowed and U1F transitions for $^{72}$Ni as a function of stellar
temperatures and densities.} \label{figure6}
\end{center}
\end{figure}

The phase space calculation for allowed and U1F transitions, as a
function of stellar temperature and density, for the neutron-rich
nickel isotope ($^{72}$Ni) is shown in Fig.~\ref{figure6}. The phase
space is calculated at selected density of 10$^{2}$ g/cm$^{3}$,
10$^{6}$ g/cm$^{3}$ and 10$^{10}$ g/cm$^{3}$ (corresponding to low,
intermediate and high stellar densities, respectively) and stellar
temperature T$_{9}$ = 0.01 - 30 given in logarithmic scale. It can
be seen from Fig.~\ref{figure6} that, for low and intermediate
stellar densities, the U1F phase space is a factor 4 bigger than the
allowed phase space at low temperatures. As stellar temperature
soars the phase space for U1F transitions is around a factor 25
bigger. For high densities the phase space is essentially zero at
small stellar temperature T$_{9} \sim$ 0.01 and increases with
increasing temperatures. At high density the U1F phase space is
around a factor 25 bigger at high temperatures. It can be further be
noted from  Fig.~\ref{figure6} that for low and intermediate stellar
densities the phase space increases by 4-8 orders of magnitude as
the stellar temperature goes from T$_{9}$ = 0.01 to 1. Moreover the
calculated phase space remains same as stellar temperature soars
from T$_{9}$ = 1 to 30. At a fixed stellar temperature, the phase
space remains the same as the core stiffens from low to intermediate
stellar density. This is because the electron distribution function
at a fixed temperature changes appreciably only once the stellar
density exceeds 10$^{7}$ g/cm$^{3}$. This happens because of an
appreciable increase in the calculated Fermi energy of the electrons
once the stellar density reaches 10$^{7}$ g/cm$^{3}$ and beyond.  As
the stellar core becomes more and more dense the phase space
decreases.  Phase space calculations of remaining isotopes namely
$^{74,76,78}$Ni shows a similar trend. When the nickel isotopes
becomes more and more neutron-rich, the phase space enhancement for
U1F transitions decreases but still is bigger than the phase space
for allowed transitions and lead to a a significant U1F contribution
to the total $\beta$--decay rates.

\begin{figure}[p]
\begin{center}
\includegraphics[width=1.0\textwidth]{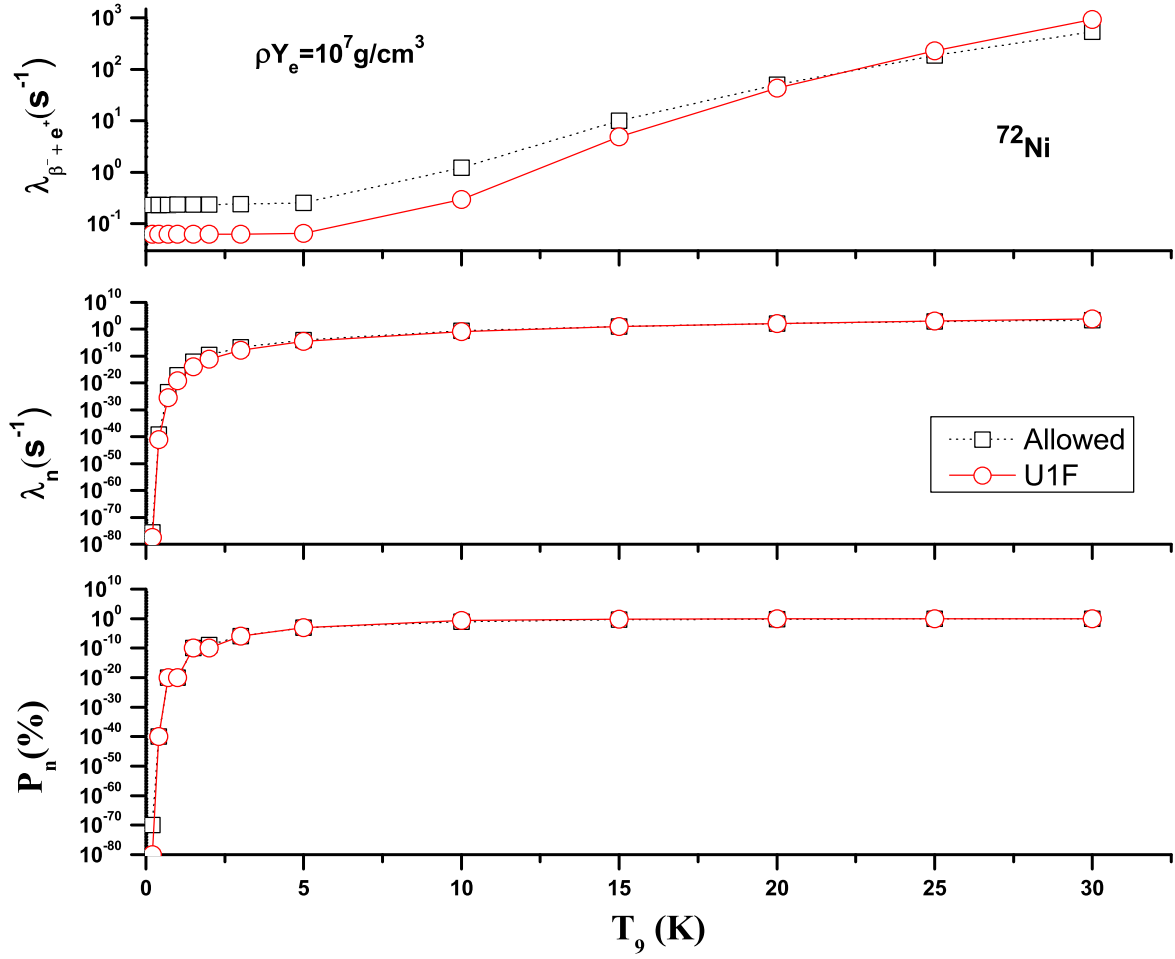}
\caption{\scriptsize Allowed (GT) and  unique first-forbidden (U1F)
$\beta$-decay $\&$ positron capture rates (upper panel) and energy
rates of neutron (middle panel) for $^{72}$Ni as a function of
temperature for selected density of 10$^{7}$g.cm$^{-3}$. Shown also
are the probabilities of $\beta$-delayed neutron emissions (bottom
panel).}\label{figure7}
\end{center}
\end{figure}

\begin{figure}[p]
\begin{center}
\includegraphics[width=1.0\textwidth]{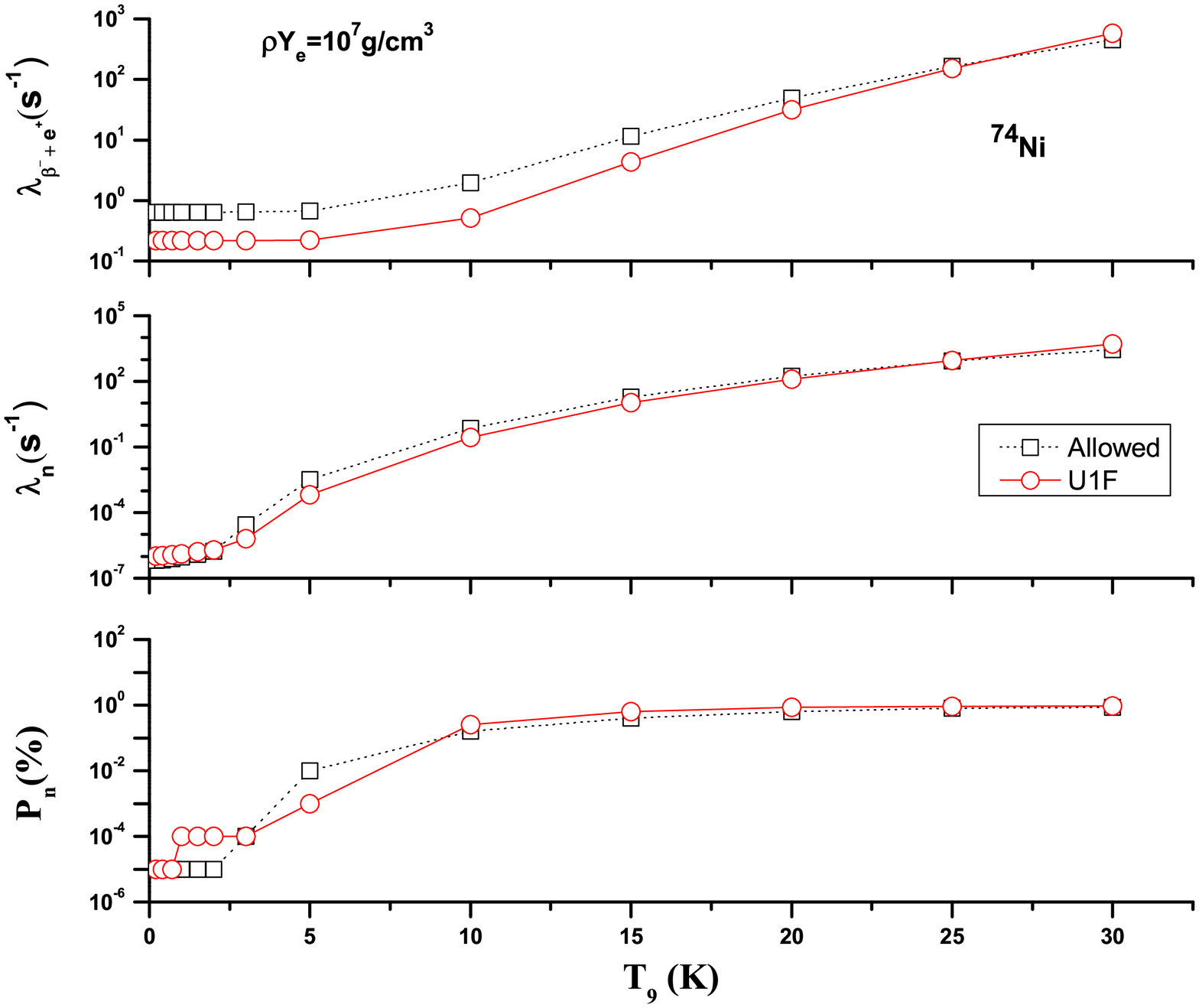}
\caption{\scriptsize Same as Fig.~\ref{figure7} but for
$^{74}$Ni.}\label{figure8}
\end{center}
\end{figure}

\begin{figure}[p]
\begin{center}
\includegraphics[width=1.0\textwidth]{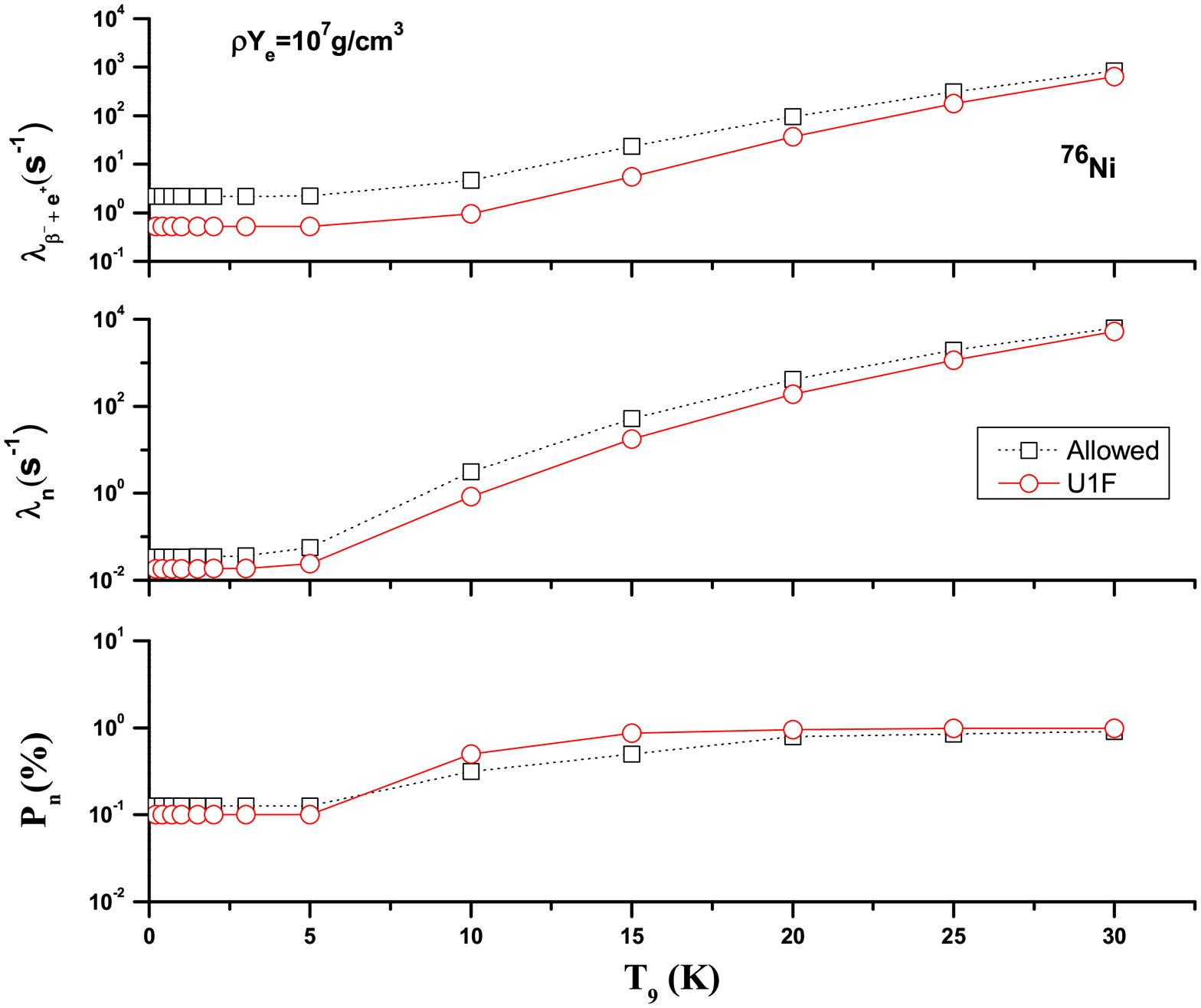}
\caption{\scriptsize Same as Fig.~\ref{figure7} but for
$^{76}$Ni.}\label{figure9}
\end{center}
\end{figure}

\begin{figure}[p]
\begin{center}
\includegraphics[width=1.0\textwidth]{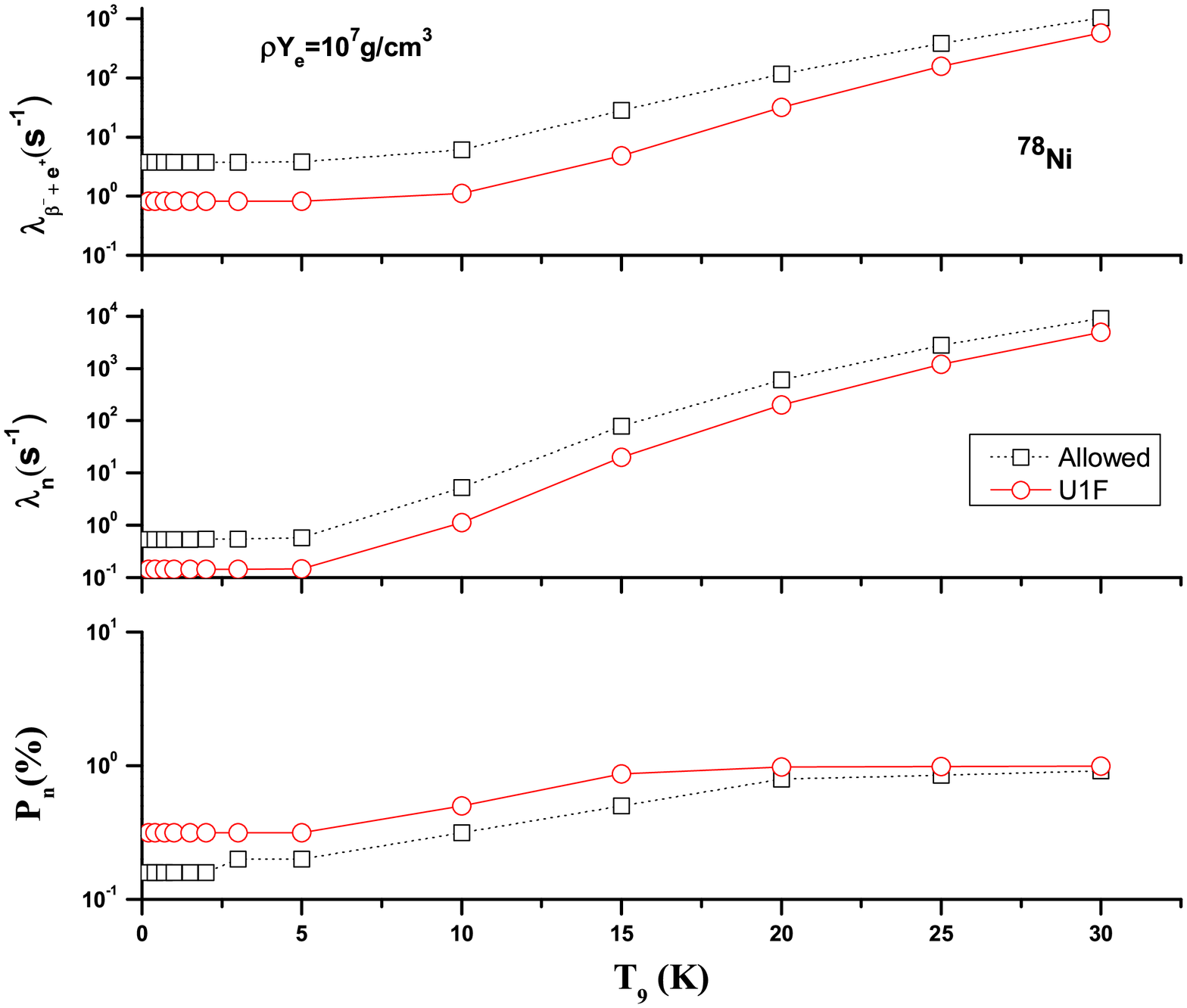}
\caption{\scriptsize Same as Fig.~\ref{figure7} but for
$^{78}$Ni.}\label{figure10}
\end{center}
\end{figure}

The stellar $\beta$-decay and positron capture rates are calculated
for nickel isotopes ($^{72,74,76,78}$Ni) in density range of
10-10$^{11}$g/cm$^{3}$ and temperature ranging from 0.01 $\leq$
T$_{9}$ $\leq$ 30. Figs.~\ref{figure7} -~\ref{figure10} show three
panels graph in which the upper panel is depicting pn-QRPA
calculated allowed and U1F $\beta$-decay+positron capture rates of
Ni isotopes at a stellar density of 10$^{7}$ g/cm$^{3}$.  It is to
be noticed that contribution from all excited states are included in
the final calculation of all rates. The allowed rates, for
intermediate density, in upper panel of Fig.~\ref{figure7}, are
roughly a factor four bigger at low temperatures and around a factor
two smaller at T$_{9}$ = 30 as compared to U1F rates. In order to
understand this behavior, one has to calculate the relative
contribution of $\beta$-decay and positron capture rates to the
total rates both for allowed GT and U1F cases. Table~\ref{ta8} shows
the relative contribution for allowed GT rates for nickel isotopes
as a function of stellar temperature at a fixed density of 10$^{7}$
g/cm$^{3}$ in the pn-QRPA(N) model. For low stellar temperatures it
is a safe assumption to neglect the positron capture rates when
compared with the $\beta$-decay rates . At high temperatures ($kT >$
1 MeV), positrons appear via electron-positron pair creation and
their capture rates exceed the competing $\beta$-decay rates by a
factor of 25 at T$_{9}$ = 30 for the case of $^{72}$Ni. Roughly same
trend is seen for  $^{74,76,78}$Ni. Table~\ref{ta9} shows a similar
comparison for the case of U1F rates. Here one notes that, at high
stellar temperatures, the contribution of positron capture rates is
much bigger to the total rates as against those of allowed GT rates
(e.g. for the case of $^{72}$Ni the positron capture rates is around
three orders of magnitude bigger than the $\beta$-decay rates at
T$_{9}$ = 30). For $^{74}$Ni, upper panel of Fig.~\ref{figure8}
shows that, at low temperatures, the allowed  rates are a factor
three bigger and as temperature soars to T$_{9}$ = 30, the U1F rates
surpass the allowed rates. For the case of $^{76}$Ni and $^{78}$Ni,
the allowed rates are bigger than the corresponding U1F rates for
all temperatures. The relative contributions of allowed and U1F
$\beta$-decay and positron capture rates as well as phase space
calculations provide the necessary explanation

The second panel in Fig.~\ref{figure7} -~\ref{figure10} depicts the
behavior of the calculated stellar rates of $\beta$-delayed neutron
emission for Ni isotopes through allowed and U1F transitions. All
rates are given in units of $MeV.s^{-1}$. The rates are calculated
for intermediate density and temperature range of 0.01 $\leq$
T$_{9}$ $\leq$ 30. In Fig.~\ref{figure7} ($^{72}$Ni) the calculated
rates of $\beta$-delayed neutron emission for allowed GT are about
two order of magnitude bigger than U1F at low temperatures. At high
temperatures the U1F rates are 3 times bigger. For the case of
$^{74}$Ni (Fig.~\ref{figure8}) the U1F neutron emission rates are
roughly twice that of allowed at low temperatures. As temperature
increases the allowed rates surpass the U1F rates. At T$_{9}$ = 30,
once again the U1F rates are twice the allowed rates. The scenario
gets interesting for the cases of $^{76}$Ni and $^{78}$Ni (see
middle panels of Fig.~\ref{figure9} and Fig.~\ref{figure10},
respectively) where the neutron emission rates from allowed
transitions are around a factor 1.5 - 4 bigger than the U1F rates
for all temperature range. The total rates are a product of phase
space and nuclear matrix elements. Whereas numerical techniques were
used for the calculation of phase space integrals, we again
re-iterate that all nuclear matrix elements were calculated in a
microscopic fashion which is a distinguishing feature of our work.

The $\beta$-delayed neutron emission probabilities ($P$$_{n}$) are
very important for a good description of both the separation
energies of the neutron (S$_{n}$) and the $\beta$ strength functions
within the Q$_{\beta}$ window. The $\beta$-delayed neutron emission
probabilities for Ni isotopes are presented in this paper for the
first time (see bottom panels of Figs.~\ref{figure7}
-~\ref{figure10}). Borzov \cite{Bor05} did comment on the $A$
dependence of the calculated $P$$_{n}$ values for Ni isotopes. We
notice a similar behavior  of increasing values of $P$$_{n}$ verses
mass number $A$ in Figs.~\ref{figure7} -~\ref{figure10}. Borzov
further noted that the increase in $\beta$-delayed neutron emission
probability for Ni isotopes with A $\leq$ 79 was entirely due to
relatively low-energy GT and FF $\beta$-decays. For all cases
($^{72,74,76,78}$Ni) the $\beta$-delayed neutron emission
probabilities due to U1F transitions are bigger than those due to
allowed GT transitions at high temperatures (T$_{9} \ge$ 10). At low
temperatures the emission probabilities are much too smaller for
$^{72,74}$Ni and becomes effective only for the neutron-richer
isotopes $^{76,78}$Ni.

\section{Conclusions}
The contribution of FF transitions to total $\beta$-decay becomes
significant for neutron-rich isotopes. We used different versions of
the pn-QRPA model using two different single-particle potentials. We
used the Woods-Saxon potential to calculate allowed GT transitions
for isotopes of nickel using the pn-QRPA(WS-SSM), pn-QRPA(WS-SPM)
and pn-QRPA(WS-DSM) models. The calculated half-lives showed
pn-QRPA(WS-DSM) to be the better model. Results of pn-QRPA(WS-SPM)
model were not very encouraging. The pn-QRPA(WS-SSM) model was later
used to calculate the FF contribution which led to a better
agreement of the calculated half-lives with the measured data. The
pn-QRPA(N) model employed the Nilsson potential and was used to
calculate allowed GT and GT+U1F half-lives. The agreement with
experimental data was the best for the pn-QRPA(N) model. The FF
inclusion improved the overall comparison of calculated terrestrial
$\beta$-decay half-lives in the pn-QRPA(WS-SSM) model. Likewise, and
more significantly,  the UIF contribution improved the pn-QRPA(N)
calculated half-lives. The DF3 + CQRPA calculation was in good
agreement with experimental data for heavier isotopes and allowed GT
calculation by M\"{o}ller and collaborators was way too big for
$^{72,74,76}$Ni but was in somewhat better agreement for $^{78}$Ni.

It was also shown that the U1F phase space has a sizeable
contribution to the total phase space at stellar temperatures and
densities. It was shown that, for a particular element, the U1F
phase space gets amplified with increasing neutron number. For the
case of pn-QRPA(N) model the U1F rates contribute roughly 20$\%$ to
the total $\beta$-decay half-lives. It is expected that contribution
of FF transition may increase further with increasing neutron
number. However there is a need to study many more neutron-rich
nuclei in order to authenticate this claim which we would like to
take as a future assignment. The microscopic calculation of U1F
$\beta$-decay rates, presented in this work, could lead to a better
understanding of the nuclear composition and $Y_{e}$ in the core
prior to collapse and collapse phase. The energy rates of
$\beta$-delayed neutrons and probability of $\beta$-delayed neutron
emissions were also calculated in stellar matter.

The reduced $\beta$-decay half-lives calculated in this work bear
consequences for nucleosynthesis problem and site-independent
$r$-process calculations. Our findings might result in speeding-up
of the $r$-matter flow relative to calculations based on half-lives
calculated from only allowed GT transitions. The effects of shorter
half-lives resulted in shifting of the third peak of the abundance
of the elements in the $r$-process toward higher mass region
\cite{Suz12}. The allowed and U1F $\beta$-decay rates on Ni isotopes
were calculated on a fine temperature-density grid, suitable for
simulation codes, and may be requested as ASCII files from the
corresponding author.

\vspace{0.5in} \textbf{Acknowledgments}

J.-U. Nabi would like to acknowledge the support of the Higher
Education Commission Pakistan through the HEC Project No. 20-3099.

\begin{table}[htbp]
\caption{Calculated GT strength for $^{72,74,76,78}$Ni using
pn-QRPA(WS-DSM, WS-SSM, WS-SPM) models.} \label{ta1}\normalsize

\hspace{0.5in}
\begin{tabular}{c|c|c|c|c|c|c}

A & E$_{j}$ &  & E$_{j}$ &  & E$_{j}$ &  \\
  &  (MeV)  & $|\langle1_{j}^{+}||M_{GT-}^{DSM}||0^{+}_{g.s}\rangle|^{2}$ &  (MeV)  & $|\langle1_{j}^{+}||M_{GT-}^{SSM}||0^{+}_{g.s}\rangle|^{2}$ &  (MeV)  & $|\langle1_{j}^{+}||M_{GT-}^{SPM}||0^{+}_{g.s}\rangle|^{2}$ \\

\hline

   & 0.01 & 6.70$\times10^{-6}$ & 0.31 & 1.40$\times10^{-3}$ & 0.88 & 4.90$\times10^{-4}$ \\
72 & 0.02 & 1.20$\times10^{-5}$ &      &                     &      &                     \\
   & 0.22 & 5.90$\times10^{-3}$ &      &                     &      &                     \\

\hline

   & 0.07 & 6.10$\times10^{-4}$ & 0.28 & 9.80$\times10^{-2}$ & 0.12 & 2.70$\times10^{-3}$ \\
74 & 0.13 & 3.30$\times10^{-3}$ & 0.43 & 3.70$\times10^{-3}$ & 1.09 & 2.30$\times10^{-3}$ \\
   & 0.57 & 1.20$\times10^{-4}$ &      &                     &      &                     \\

\hline

   & 0.50 & 1.20$\times10^{-4}$ & 0.32 & 1.04$\times10^{-1}$ & 0.28 & 4.50$\times10^{-2}$ \\
   & 0.60 & 1.40$\times10^{-4}$ & 1.44 & 6.50$\times10^{-2}$ & 0.78 & 2.30$\times10^{-2}$ \\
76 & 0.75 & 2.30$\times10^{-4}$ & 1.57 & 1.40$\times10^{-3}$ & 1.91 & 8.20$\times10^{-3}$ \\
   & 0.85 & 2.90$\times10^{-4}$ &      &                     &      &                     \\
   & 0.86 & 3.00$\times10^{-4}$ &      &                     &      &                     \\

\hline

   & 1.00 & 9.90$\times10^{-5}$ & 0.81 & 3.70$\times10^{-2}$ & 0.49 & 5.30$\times10^{-4}$ \\
78 & 1.40 & 2.50$\times10^{-4}$ & 0.91 & 2.07$\times10^{-3}$ & 0.89 & 1.30$\times10^{-2}$ \\
   & 1.70 & 5.50$\times10^{-4}$ &      &                     &      &                    \\

\end{tabular}
\end{table}

\begin{table}[htbp]
\caption{Calculated GT strengths (rank0, rank1 and rank2
transitions) for $^{72,74,76,78}$Ni using  the pn-QRPA (WS-SSM)
model.} \label{ta2}\normalsize \hspace{0.7in}
\begin{tabular}{c|c|c|c|c|c|c}

A & E$_{j}$ &  & E$_{j}$ &  & E$_{j}$ &  \\
  &  (MeV)  & $|\langle0_{j}^{-}||M_{\beta-}^{0}||0^{+}\rangle|^{2}$ &  (MeV)  & $|\langle1_{j}^{+}||M_{\beta-}^{1}||0^{+}\rangle|^{2}$ &  (MeV)  & $|\langle2_{j}^{+}||M_{\beta-}^{2}||0^{+}\rangle|^{2}$ \\

\hline

72 & 2.01 & 1.80$\times10^{-4}$ & 0.95 & 3.4$\times10^{-4}$ & 2.15 & 2.40$\times10^{-2}$ \\

74 & 1.76 & 2.30$\times10^{-4}$ & 0.18 & 3.3$\times10^{-4}$ & 0.03 & 8.20$\times10^{-3}$  \\

76 & 0.19 & 3.10$\times10^{-4}$ & 1.47 & 9.50$\times10^{-4}$ & 1.49 & 6.70$\times10^{-3}$ \\

78 & 1.63 & 5.90$\times10^{-4}$ & 0.26 & 3.10$\times10^{-4}$ & 0.21 & 6.80$\times10^{-3}$ \\
\end{tabular}
\end{table}

\begin{table}[htbp]
\caption{Deformation parameters used in the pn-QRPA(N) and
pn-QRPA(WS-DSM) calculations.} \label{ta3} \normalsize\hspace{3.0in}
\begin{tabular}{c|c}

A & $\delta$ \\

\hline
72 & 0.00896 \\
74 & 0.01583 \\
76 & 0.01037 \\
78 & 0.00340 \\
\end{tabular}
\end{table}

\begin{table}[htbp]
\caption{Allowed GT $\beta$-decay half-lives (in seconds) for Ni
isotopes calculated using the pn-QRPA(N) and pn-QRPA(WS-DSM, WS-SSM,
WS-SPM) models, in comparison with experimental data \cite{Aud12}
and those by Ref. \cite{Moe97}.} \label{ta6} \footnotesize
\hspace{1.3in}
\begin{tabular}{c|c|c|c|c|c|c}

   &     & pn-QRPA &  pn-QRPA & pn-QRPA  & pn-QRPA  &            \\
A  & Exp & (N)     & (WS-DSM) & (WS-SSM) & (WS-SPM) & M\"{o}ller \\
   &     & (GT)    & (GT)     & (GT)     & (GT)     & (GT)       \\

\hline
72 & 1.57 & 2.50 & 1.04 & 1.12 & 12.6 & 42.7 \\
74 & 0.68 & 0.98 & 0.71 & 0.35 & 1.23 & 26.8 \\
76 & 0.24 & 0.29 & 0.20 & 1.07 & 0.18 & 3.07 \\
78 & 0.14 & 0.17 & 0.11 & 0.16 & 0.12 & 0.22 \\
\end{tabular}
\end{table}

\begin{table}[htbp]
\caption{Total $\beta$-decay half-lives (in seconds) for Ni isotopes
calculated using the pn-QRPA(N) and pn-QRPA(WS-SSM) models for
allowed plus  first-forbidden  transitions, in comparison with
experimental data \cite{Aud12} and the DF3 + CQRPA \cite{Bor05}
calculation.} \label{ta7} \footnotesize \hspace{1.0in}
\begin{tabular}{c|c|c|c|c|c|c|c}

   &     & pn-QRPA         &  pn-QRPA        & pn-QRPA             & pn-QRPA         &                 \\
A  & Exp & (N)             & (WS-SSM)        & (WS-SSM)            & (WS-SSM)        & DF3+CQRPA       \\
   &     & (GT+rank$_{2}$) & (GT+rank$_{2}$) & (GT+rank$_{0,1,2}$) & (GT+rank$_{0}$) & (GT+rank$_{0}$) \\

\hline
72 & 1.57 & 2.01 & 1.12 & 1.09 & 1.11 & 1.33 \\
74 & 0.68 & 0.74 & 0.34 & 0.32 & 0.35 & 0.53 \\
76 & 0.24 & 0.24 & 1.01 & 0.89 & 1.03 & 0.26 \\
78 & 0.14 & 0.14 & 0.15 & 0.15 & 0.15 & 0.13 \\
\end{tabular}
\end{table}

\begin{table}[htbp]
\caption{Calculated $\beta^{-}$-decay rates for allowed GT
transitions (in units of s$^{-1}$) and ratio of $\beta^{-}$-decay to
positron capture rates on $^{72,74,76,78}$Ni at stellar density of
10$^{7}$g/cm$^{3}$ in the pn-QRPA(N) model. T$_{9}$ represents the
temperature in 10$^{9}$K.} \label{ta8} \normalsize\hspace{0.3in}
\begin{tabular}{c|c|c|c|c|c|c|c|c}

 & \multicolumn{2}{|c|}{$^{72}$Ni} & \multicolumn{2}{|c|}{$^{74}$Ni} & \multicolumn{2}{|c|}{$^{76}$Ni} & \multicolumn{2}{|c}{$^{78}$Ni} \\
\cline{2-9}
T$_{9}$ & $\lambda_{\beta^{-}}$ & R($\beta^{-}/e^{+}$) & $\lambda_{\beta^{-}}$ & R($\beta^{-}/e^{+}$) & $\lambda_{\beta^{-}}$ & R($\beta^{-}/e^{+}$) & $\lambda_{\beta^{-}}$ & R($\beta^{-}/e^{+}$) \\

\hline
1    & 2.3$\times10^{-1}$ & 1.4$\times10^{11}$ & 6.3$\times10^{-1}$ & 2.8$\times10^{11}$ & 2.1$\times10^{0}$ & 5.0$\times10^{11}$  & 3.7$\times10^{0}$ & 7.7$\times10^{11}$  \\
3    & 2.4$\times10^{-1}$ & 8.2$\times10^{3}$  & 6.5$\times10^{-1}$ & 1.7$\times10^{4}$  & 2.2$\times10^{0}$ & 3.2$\times10^{4}$   & 3.7$\times10^{0}$ & 5.0$\times10^{4}$   \\
5    & 2.6$\times10^{-1}$ & 8.6$\times10^{1}$  & 6.7$\times10^{-1}$ & 1.9$\times10^{2}$  & 2.3$\times10^{0}$ & 3.6$\times10^{2}$   & 3.8$\times10^{0}$ & 5.6$\times10^{2}$   \\
10   & 8.7$\times10^{-1}$ & 2.4$\times10^{0}$  & 1.6$\times10^{0}$  & 4.3$\times10^{0}$  & 4.1$\times10^{0}$ & 7.3$\times10^{1}$   & 5.5$\times10^{0}$ & 1.1$\times10^{1}$   \\
30   & 2.1$\times10^{1}$  & 4.0$\times10^{-2}$ & 2.6$\times10^{1}$  & 6.0$\times10^{-2}$ & 7.2$\times10^{1}$ & 9.3$\times10^{-2}$  & 1.3$\times10^{2}$ & 1.4$\times10^{-1}$  \\
\end{tabular}
\end{table}

\begin{table}[htbp]
\caption{Same as Table~\ref{ta8} but for U1F transitions.}
\label{ta9} \normalsize\hspace{0.3in}
\begin{tabular}{c|c|c|c|c|c|c|c|c}

 & \multicolumn{2}{|c|}{$^{72}$Ni} & \multicolumn{2}{|c|}{$^{74}$Ni} & \multicolumn{2}{|c|}{$^{76}$Ni} & \multicolumn{2}{|c}{$^{78}$Ni} \\
\cline{2-9}
T$_{9}$ & $\lambda_{\beta^{-}}$ & R($\beta^{-}/e^{+}$) & $\lambda_{\beta^{-}}$ & R($\beta^{-}/e^{+}$) & $\lambda_{\beta^{-}}$ & R($\beta^{-}/e^{+}$) & $\lambda_{\beta^{-}}$ & R($\beta^{-}/e^{+}$) \\

\hline
1    & 6.3$\times10^{-2}$ & 5.8$\times10^{11}$ & 2.1$\times10^{-1}$ & 1.7$\times10^{12}$ & 5.1$\times10^{-1}$ & 2.8$\times10^{12}$ & 8.2$\times10^{-1}$ & 4.1$\times10^{12}$  \\
3    & 6.4$\times10^{-2}$ & 1.5$\times10^{4}$  & 2.1$\times10^{-1}$ & 4.6$\times10^{4}$  & 5.1$\times10^{-1}$ & 8.1$\times10^{4}$  & 8.2$\times10^{-1}$ & 1.2$\times10^{5}$   \\
5    & 6.5$\times10^{-2}$ & 9.8$\times10^{1}$  & 2.2$\times10^{-1}$ & 3.0$\times10^{2}$  & 5.2$\times10^{-1}$ & 5.6$\times10^{2}$  & 8.2$\times10^{-1}$ & 8.7$\times10^{2}$   \\
10   & 1.0$\times10^{-1}$ & 5.2$\times10^{-1}$ & 3.5$\times10^{-1}$ & 2.0$\times10^{0}$  & 7.7$\times10^{-1}$ & 4.4$\times10^{0}$  & 9.7$\times10^{-1}$ & 7.1$\times10^{0}$   \\
30   & 5.0$\times10^{-1}$ & 5.4$\times10^{-4}$ & 1.6$\times10^{0}$  & 2.8$\times10^{-3}$ & 5.3$\times10^{0}$  & 8.4$\times10^{-3}$ & 1.1$\times10^{1}$  & 1.9$\times10^{-2}$  \\
\end{tabular}
\end{table}




\clearpage
\clearpage
\clearpage
\clearpage
\clearpage
\clearpage
\clearpage
\clearpage
\clearpage

\end{document}